\documentclass{article}
\usepackage{spconf,amsmath,graphicx}
\usepackage{xcolor}
\usepackage{color, colortbl}
\usepackage{multirow}
\usepackage{tablefootnote}
\usepackage{comment}
\usepackage{float}
\usepackage{refcount}
\usepackage{subcaption}
\usepackage{paralist}
\usepackage{amsfonts}
\usepackage{amssymb}
\usepackage{booktabs}

\definecolor{LightCyan}{rgb}{0.88,1,1}

\title{INPUT-INDEPENDENT ATTENTION WEIGHTS ARE EXPRESSIVE ENOUGH:\\
A STUDY OF ATTENTION IN SELF-SUPERVISED AUDIO TRANSFORMERS}
\name{Tsung-Han Wu, Chun-Cheng Hsieh, Yen-Hao Chen, Po-Han Chi,  Hung-yi Lee}
\address{
  National Taiwan University,\\
  College of Electrical Engineering and Computer Science
  }
\begin{document}
%
\maketitle
\begin{abstract}
In this paper, we seek solutions for reducing the computation complexity of transformer-based models for speech representation learning. We evaluate 10 attention algorithms; then, we pre-train the transformer-based model with those attention algorithms in a self-supervised fashion and treat them as feature extractors on downstream tasks, including phoneme classification and speaker classification. With the assistance of t-SNE, PCA and some observation, the attention weights in self-supervised audio transformers can be categorized into four general cases. Based on these cases and some analyses, we are able to use a specific set of attention weights to initialize the model. Our approach shows comparable performance to the typical self-attention yet requires $20\%$ less time in both training and inference.


\end{abstract}

\section{Background}
Recently, self-supervised training is used widely in representation learning due to its good ability of generalization.~\cite{devlin2018bert,radford2019language,Schneider2019,jiang2019improving,oord2018representation,jing2020self} Generally speaking, there are two training phases for  typical self-supervised learning: pre-training and fine-tuning. In the pre-training phase, models are usually trained with a large amount of unlabeled data. As for the fine-tuning phase, the pre-trained models are fine-tuned on some downstream tasks with limited labeled data.

In the Natural Language Processing (NLP) domain, BERT~\cite{devlin2018bert} is the most popular feature extractor using self-supervised training. BERT makes use of transformers, and the pre-training technique BERT uses is called Masked Language Model (MLM). The key idea of MLM is that given some input sequences with partially masked tokens, the model should be able to predict those masked tokens.
Because of the great success of BERT in the NLP domain, Masked Audio Model (MAM)~\cite{liu2019mockingjay} is proposed to pre-train robust audio representations. In summary, MAM is similar to MLM; it masks some portions of the input audio signals, then makes models to reconstruct those masked part. Identical to BERT, models trained with MAM are repurposed as feature extractors on different downstream tasks. 
\label{sec:background}

\section{Introduction}
\label{sec:intro}
Transformers~\cite{vaswani2017attention} have become the most powerful network architecture in many fields, including speech, natural language processing (NLP), computer vision (CV), etc. 
Its outstanding performance is based not only on capturing long-term dependencies of input sequence but also on the remarkable training efficiency. 
Among all transformer-based models, BERT~\cite{devlin2018bert} is probably the most famous one. It can learn strong language representations from unlabeled text. 
The idea of training transformer-based models on unlabeled audio data has also been widely studied~\cite{song2019speech, jiang2019improving, baevski2019effectiveness, schneider2019wav2vec, pascual2019learning}. 

Nonetheless, transformer-based models are usually based on pretty high computational cost because of their self-attention mechanism. 
Despite the effectiveness of self-attention~\cite{lin2017structured},
it suffers severely from both quadratic memory and computation requirements with respect to the input sequence length. 
The input sentences in NLP come across this problem, not to mention those much longer input sequence in speech. 
Therefore, several new attention mechanisms are proposed to reduce the time complexity.

In this work we summarize variants of attention mechanisms for transformers.
Sparse Transformers~\cite{child2019generating} proposes two attention masks which have lower time complexity and satisfactory performance.
Routing Transformers~\cite{roy2020efficient} uses K-means clustering to determine the candidates for each query in self-attention, while Reformer~\cite{kitaev2020reformer} tries to address this issue by introducing the hashing algorithm.
 
We also introduce \textit{locality-sensitive hashing (LSH)}, which can solve the \textit{Maximum Inner Product Search} (MIPS) problem efficiently~\cite{huang2018accurate}; the self-attention mechanism in transformers can be regarded as a MIPS problem.
Asymmetric LSH (ALSH)~\cite{shrivastava2014asymmetric} provides the first provably sublinear time hashing algorithm for MIPS, and they successfully prove that there does not exist any LSH family for MIPS. Subsequently, the same authors propose an improved ALSH algorithm~\cite{shrivastava2014improved}, which reaches better performance. XBOX~\cite{bachrach2014speeding} is another ALSH algorithm. At the same time, Simple LSH~\cite{neyshabur2014symmetric} argues that neither LSH nor ALSH can deal with MIPS well and proposes two stronger LSH/ALSH algorithms. Then, Query Normalized First (QNF)~\cite{huang2018accurate} is proposed, which is a method that can be used together with those previously mentioned LSH algorithms and shows superior empirical performance.

Furthermore, aside from those \textit{crafted} attention masks and  hashing methods, there still exist some other algorithms that can reduce the time complexity of self-attention. 
Adaptive Attention~\cite{sukhbaatar2019adaptive} is a self-attention mechanism that can learn its optimal attention span, allowing models to deal with longer sequences.
Fixed attention patterns~\cite{raganato2020fixed} is another method meant to boost the performance in low-resource scenarios.
Longformer~\cite{beltagy2020longformer} introduces the attention mechanism scaling linearly with the input length. This modification makes it easy to process longer input sequences. Also, they bring the idea of dilated attention analogous to dilated CNNs~\cite{oord2016wavenet}. Lite Transformer~\cite{wu2020lite} uses CNNs along with the original self-attention to accelerate and boost the performance. 
Last but not least, SYNTHESIZER~\cite{tay2020SYNTHESIZER} proposes two simple yet effective ways to generate \textit{attention weights} directly without token-token interactions. Note that SYNTHESIZER differs from those LSH/ALSH attentions and Sparse attentions, all of which try to generate \textit{attention masks}  from token-token interactions. The difference between \textit{attention masks} and \textit{attention weights} can refer to Figure~\ref{fig:syn_transformer}. This modification accelerates both the training and inference speed drastically. 

Among these attentions, some of them have already been realized on NLP or CV tasks, whereas the others are merely examined with either mathematical theories or some simple experiments. In this work, the analyses of attention algorithms and our proposed method are all based on self-supervised audio transformers~\footnote{We will publish source code on Github in the camera-ready version.}.

\begin{figure}[t]
\centering
\begin{subfigure}[t]{.24\columnwidth}
  \includegraphics[width=.99\textwidth]{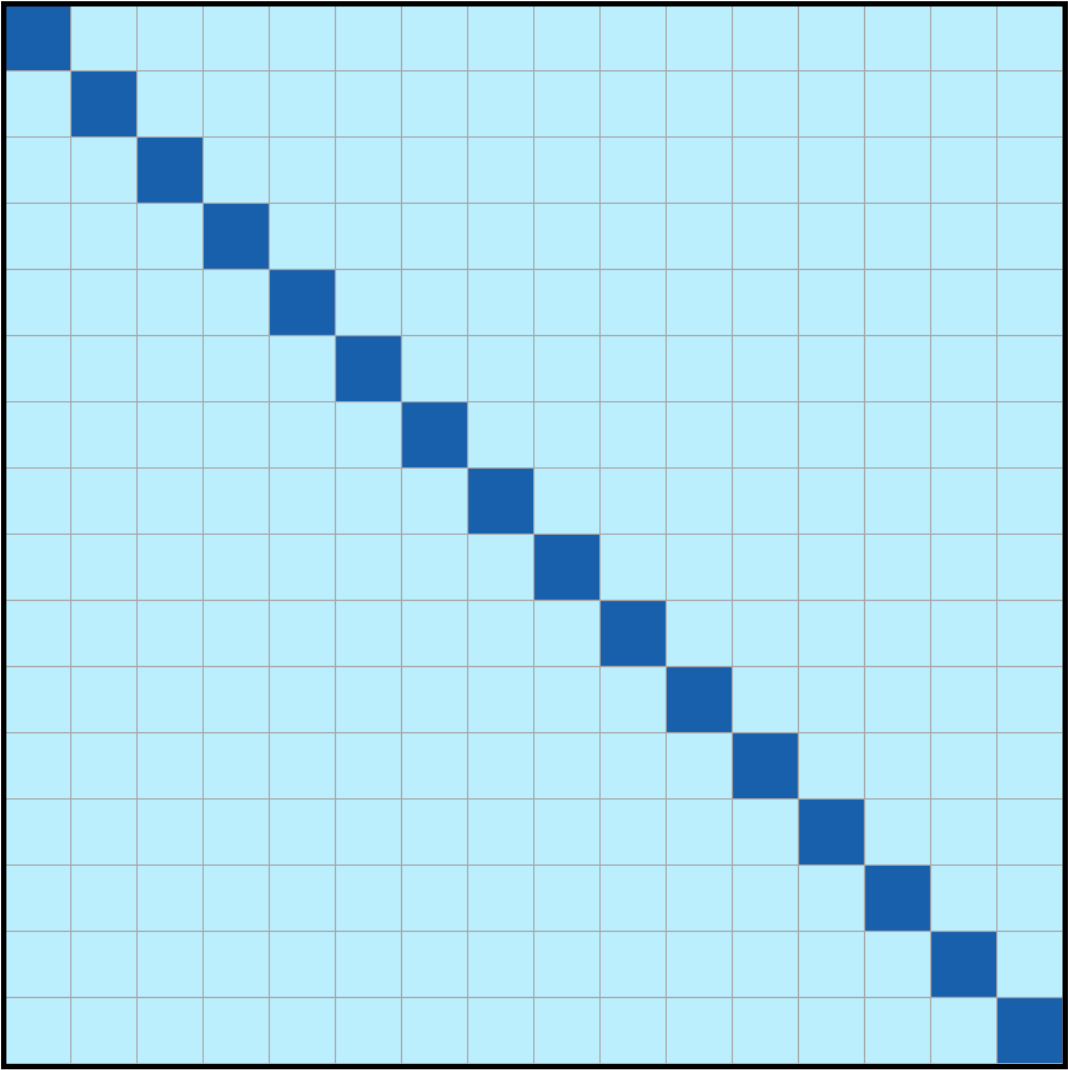}
  \caption{Full $n^2$ attenion}
  \label{fig:attenion_full}
\end{subfigure}
\begin{subfigure}[t]{.24\columnwidth}
  \includegraphics[width=.99\textwidth]{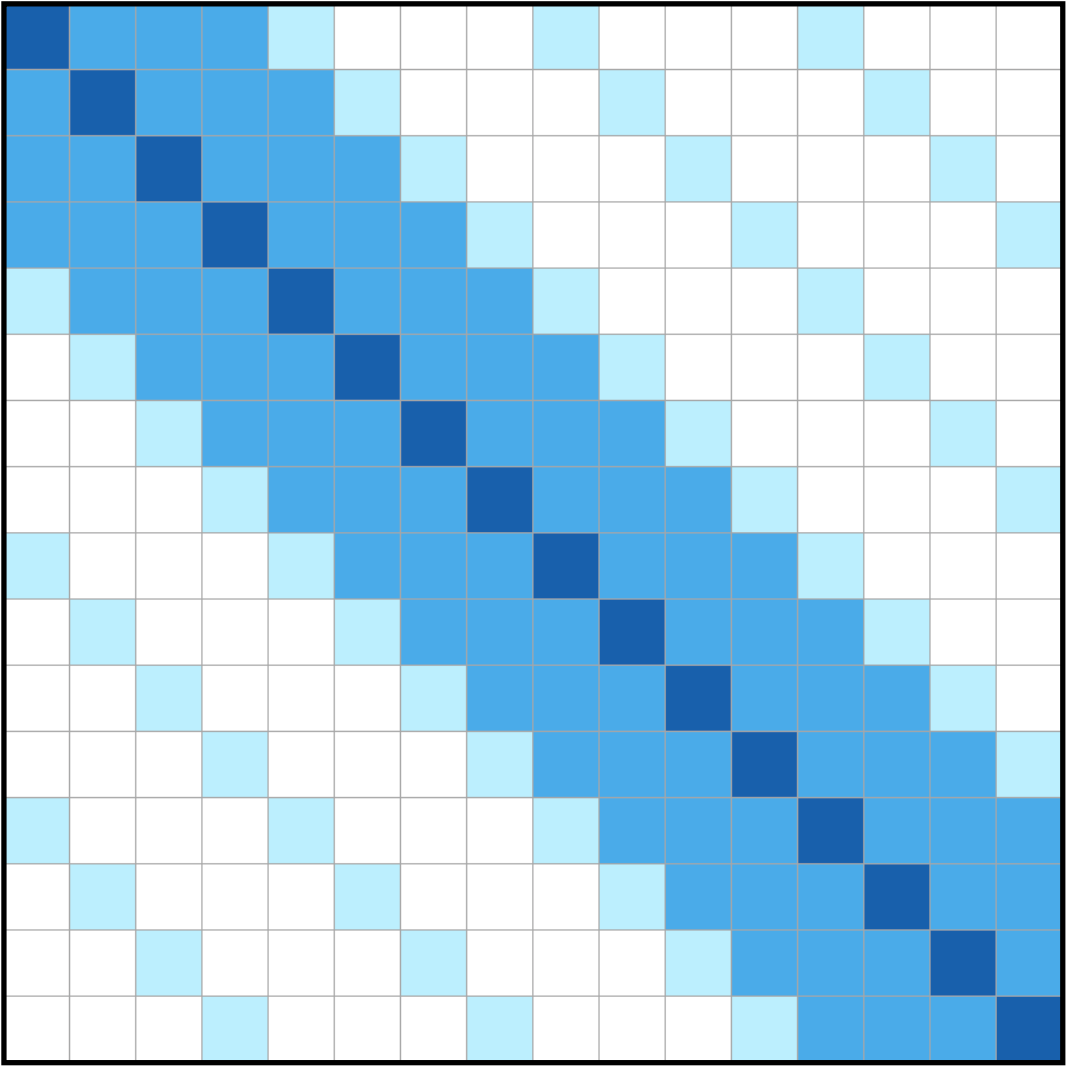}  
  \caption{Sparse attention (strided)}
  \label{fig:attention_strided}
\end{subfigure}
\begin{subfigure}[t]{.24\columnwidth}
  \includegraphics[width=.99\linewidth]{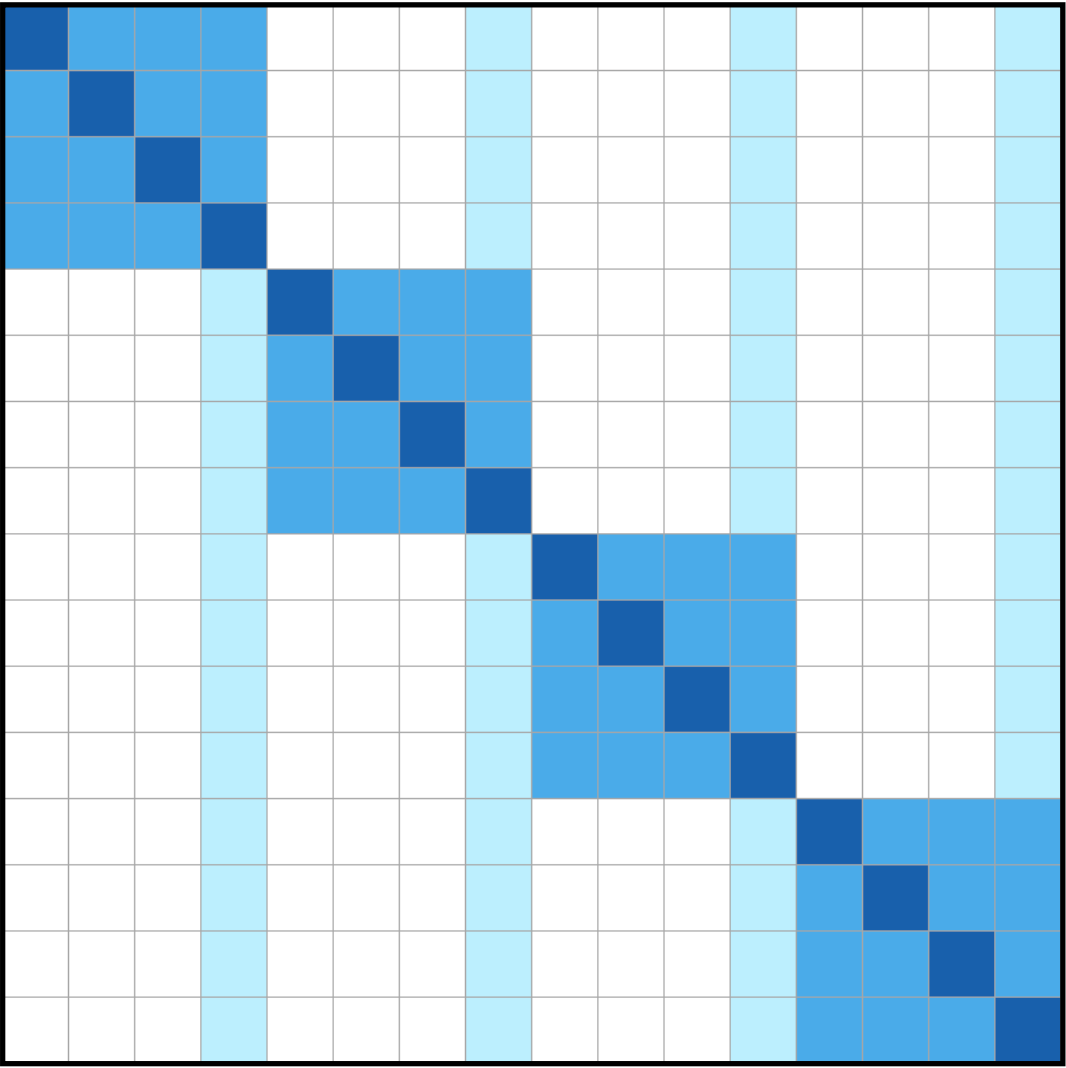}  
  \caption{Sparse attention (fixed)}
  \label{fig:attention_fixed}
\end{subfigure}
\begin{subfigure}[t]{.24\columnwidth}

  \includegraphics[width=.99\linewidth]{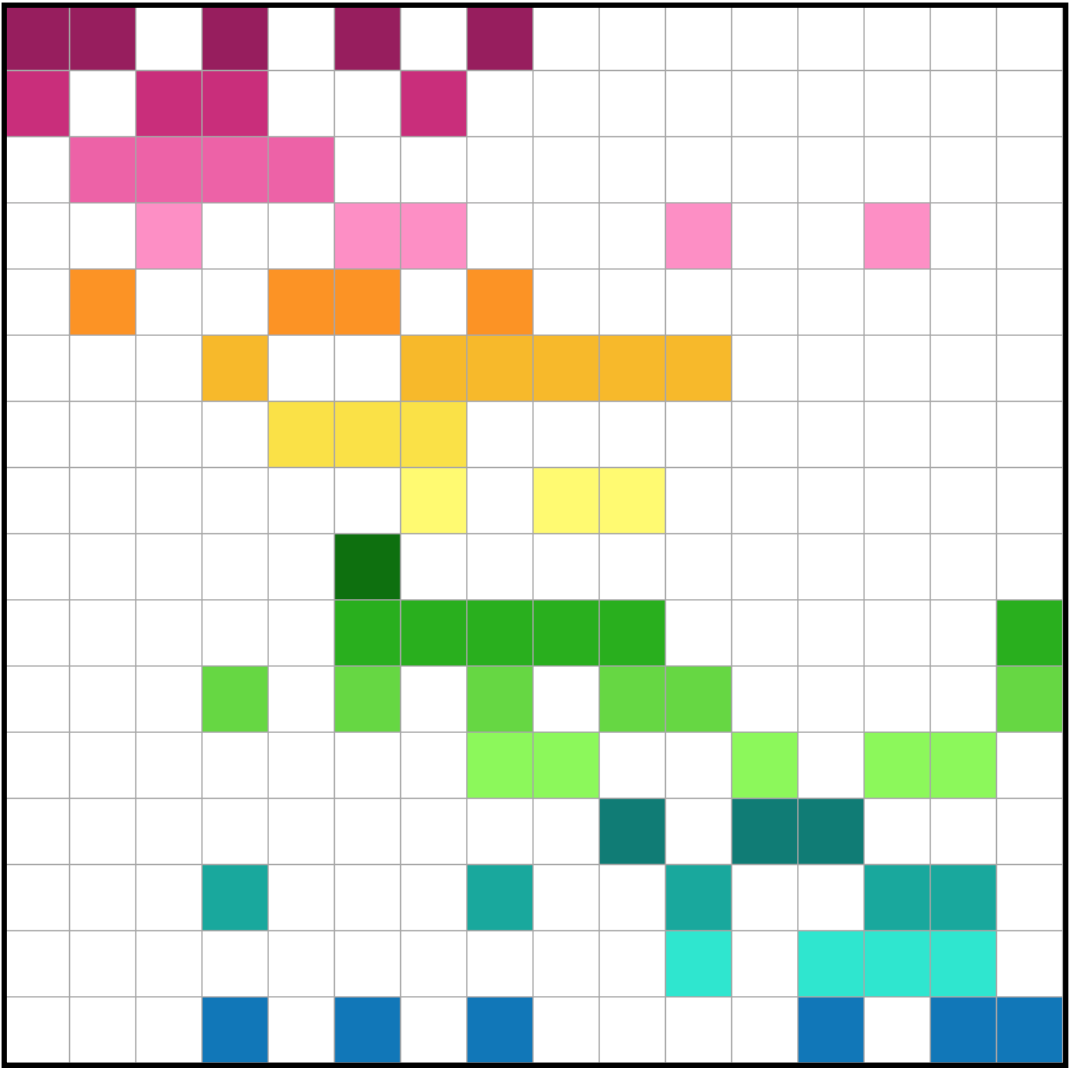}  
  \caption{LSH/ALSH attention}
  \label{fig:attention_hash}
\end{subfigure}

\caption{Comparison between different attention mechanism. These figures show the connections between the queries (columns) and the keys (rows). The colored squares represent the candidates that should be attended to in self-attention. In (a), dark blue squares represent queries whereas light blue squares represent the key indices attended to. In (b) and (c), dark blue squares are still queries, the medium light blue squares represent the key indices being attended to by some attention heads, and the light blue squares represent the key indices being attended to by the other attention heads. Lastly, in (d), for each query, LSH/ALSH algorithms will determine the key indices to be attended to.}
\label{fig:attention}
\end{figure}

In this paper, we have the following key contributions:
\begin{enumerate}
    \item We survey and implement these attention algorithms respectively, trying to figure out their efficiency and effectiveness on self-supervised transformer-based models. Figure~\ref{fig:attention} shows different types of attention algorithms we implement, while Table~\ref{tab:summary} summarizes all of the attention algorithms along with their theoretical and practical running time.  
    \item The analyses of attention weights bring a fresh viewpoint on attention mechanism in audio transformers, providing a descriptive basis for additional research. 
    \item We propose an attention mechanism, yielding the competitive performance yet with a reduction of time in both training and inference.
\end{enumerate}

\begin{figure*}[t]
\centering
\begin{subfigure}[t]{0.99\textwidth}
  \includegraphics[width=.99\textwidth]{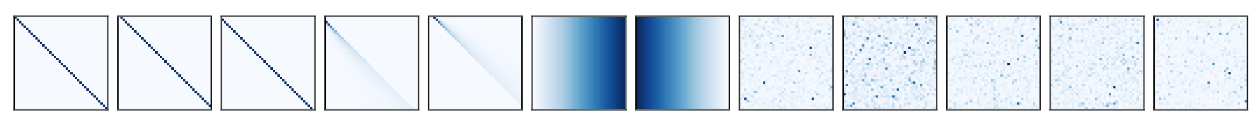}
  \caption{Attention weights we used to initialize our model. The five leftmost weights are diagonal, shifting diagonal left one step, two steps, and shifting diagonal right one step, two steps. The five rightmost weights are initialized with small random numbers. As for the middle two weights, we initialize them using the numbers increasing from left to right and decreasing from left to right respectively.}
  \label{fig:attenion_weight_init}
\end{subfigure}

\begin{subfigure}[t]{0.99\textwidth}
  \includegraphics[width=.99\textwidth]{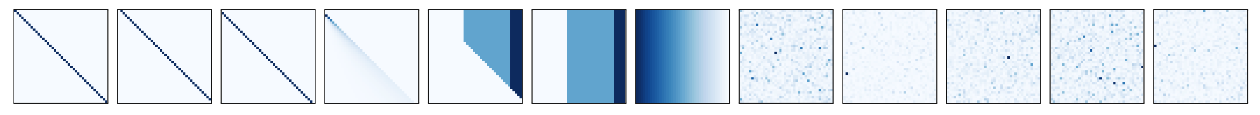}
  \caption{Attention weights learned by our model after pre-training.}
  \label{fig:attenion_weight}
\end{subfigure}
\caption{There are 12 attention heads in our models.}
\end{figure*}

\section{Methodology}
\label{sec:method}
The study was designed to answer the following research questions: Are those new attention algorithms helpful for audio transformers? If so, is there any better architecture or algorithm? We investigated these questions by applying the proposed attention algorithms on a 6-layer audio transformer. 

\subsection{Self-attention}
\label{sec:self-attention}
Each transformer layer takes an inptut vector sequence $X = x_1, ..., x_L$, and $X \in \mathbb{R}^{L \times D}$, where $L$ is \textit{input sequence length} and $D$ stands for the hidden dimension. 
Next, three different linear layers project the input vector sequence $X$ to its corresponding query matrix ($Q \in {L \times D}$), key matrix ($K \in {L \times D}$), and value matrix ($V \in {L \times D}$), respectively. 
Each vector $x_i$ in $X$ has a query vector $q_i$ ($i$-th row of $Q$), key vector $k_i$ ($i$-th row of $K$), and value vector $v_i$ ($i$-th row of $V$).
For standard single-head attention~\cite{vaswani2017attention}, the attention weight matrix $A\in \mathbb{R}^{L \times L}$ is generated by multiplying query $Q$ and key $K$ transpose.
An element in $A[i,j]$ is computed as below.
\begin{equation}
A[i,j] = \frac{q_i \dot k_j}{\sqrt{D}}.
\label{eq:A}
\end{equation}
For multi-head attention, $A_1,...,A_H$ are geneated, where  $H$ stands for the number of attention heads.

Although multi-head attention is powerful, it requires an enormous amount of computations. This situation deteriorates as input sequences become longer and longer. The primitive idea is to restrict the number of keys attended to for each query. More specifically, the number of keys attended to should not have a strong positive correlation with the input length. This problem can be solved efficiently by locality-sensitive hashing (LSH). In the following subsections, we will elaborate on all LSH algorithms we implement in this paper.



\subsection{Locality-sensitive hashing (LSH)}
\label{sec:LSH}
In (\ref{eq:A}), the dot-products of all pairs of $q_i$ and $k_j$ have to be computed. 
The basic idea of LSH is to quickly identify pairs of $q_i$ and $k_j$ leading to large enough dot-products. Only the dot-products of the identified pairs have to be computed, while the rest are directly set to zero. 

In general, there exist two different transformations $S(x)$ and $R(x)$ in \textit{asymmetric} LSH (ALSH) and exactly one transformation $S(x)$ in \textit{symmetric} LSH (LSH).
Both  $S(x)$ and $R(x)$ takes a vector $x$ as input and outputs another vector.
In the asymmetric case, query vectors $q$ and key vectors $k$ are encoded with different transformations $S(x)$ and $R(x)$ respectively\footnote{$x$ can be $q$  or $k$.}, while we encode $q$ and $k$  with the same transformation $S(x)$ in the symmetric one. 
Then, we define the hash function as:
\begin{align}
    h^{Sign}\left(x\right)=sign\left(a^\top x\right),
\end{align}
where $a$ is a random vector with $a_i\sim N\left(0,1\right)$. For ALSH, if $h^{Sign}\left(S\left(q\right)\right)=h^{Sign}\left(R\left(k\right)\right)$, query $q$ will attend to this key $k$; otherwise, we take no action. 
For LSH, query $q$ attend to the key $k$ only if $h^{Sign}\left(S\left(q\right)\right)=h^{Sign}\left(S\left(k\right)\right)$. Specifically, we define the hyperparameter $C$ to control the number of keys being attended to. 
That is, we choose only $C$ keys instead of all the keys that meet the condition  $h^{Sign}\left(S\left(q\right)\right)=h^{Sign}\left(S\left(k\right)\right)$.
Here we directly choose the top $C$ values of keys:
\begin{align}
    \begin{cases}
        a^\top k,\ h^{Sign}\left(q\right)\geq0\\
        -a^\top k,\ h^{Sign}\left(q\right)<0.
    \end{cases}
\end{align}

Due to the space limitation, we briefly list the formulation of the LSH algorithm implemented in this paper.
Please refer to the original paper for further explanation of each algorithm.
Additionally, all queries and keys are normalized with some transformations $T(x)$ before the hashing functions. Note that for different LSH algorithms, the normalization methods may differ, but the normalization is not explicitly formulated in the following description for simplicity.

\subsubsection{Sign-ALSH}
\label{sec:sign-alsh}
For Sign-ALSH~\cite{shrivastava2014improved}, there are two hyperparameters $U$ and $m$ used for normalization. Let $M\triangleq\max_{x\in\mathcal{X}}\left\|x\right\|_2$ and define the transformations $T:\mathbb{R}^D\rightarrow\mathbb{R}^D$
\begin{align}
    T\left(x\right)=Ux/M,
\end{align}
and $S,R:\mathbb{R}^D\rightarrow\mathbb{R}^{D+m}$:
\begin{align}
    S\left(x\right)&=\left[x;0;0;...;0\right],\\
    R\left(x\right)&=\left[x;\frac{1}{2}-\left\|x\right\|^2_2;\frac{1}{2}-\left\|x\right\|^4_2;...;\frac{1}{2}-\left\|x\right\|^{2^m}_2\right].
\end{align}
\subsubsection{XBOX}
XBOX~\cite{bachrach2014speeding} is an \textit{asymmetric} LSH; neither normalization nor hyperparameters is in XBOX. Then, define transformations $S,R:\mathbb{R}^D\rightarrow\mathbb{R}^{D+1}$:
\begin{align}
    S\left(x\right)&=\left[x;0\right],\\
    R\left(x\right)&=\left[x;\sqrt{M^2-\left\|x\right\|^2_2}\right],
\end{align}
where $M\triangleq\max_{x\in\mathcal{X}}\left\|x\right\|_2$.
\subsubsection{Simple LSH \& Simple ALSH}
There is no hyperparameter in Simple LSH and Simple ALSH~\cite{neyshabur2014symmetric}. As for the normalization, let $M\triangleq\max_{x\in\mathcal{X}}\|x\|_2$, then define the transformation $T:\mathbb{R}^D\rightarrow\mathbb{R}^D$: 
\begin{align}
    T(x)=\frac{x}{M}.
\end{align}
For Simple LSH, define the transformation $S:\mathbb{R}^D\rightarrow\mathbb{R}^{D+1}$:
\begin{align}
    S\left(x\right)=\left[x;\sqrt{1-\|x\|^2_2}\right],
\end{align}
whereas for Simple ALSH, define two transformations $S,R:\mathbb{R}^D\rightarrow\mathbb{R}^{D+2}$:
\begin{align}
    S\left(x\right)&=\left[x;0;\sqrt{1-\|x\|^2_2}\right],\\
    R\left(x\right)&=\left[x;\sqrt{1-\|x\|^2_2};0\right].
\end{align}
\subsubsection{Query Normalized First (QNF)}
In QNF~\cite{huang2018accurate}, let $M\triangleq\max_{x\in\mathcal{X}}\left\|x\right\|_2$ and define transformations $S,R:\mathbb{R}^D\rightarrow\mathbb{R}^{D+1}$:
\begin{align}
    S\left(x\right)&=\left[\lambda x;0\right],{\rm where}\ \lambda=\frac{M}{\left\|x\right\|_2},\\
    R\left(x\right)&=\left[x;\sqrt{M^2-\|x\|^2_2}\right].
\end{align}

\subsection{Sparse attention} \label{subsec:sparse}
Instead of using algorithms to determine which keys should be attended to, Sparse Attention~\cite{child2019generating} merely \textit{crafts} attention masks $M \in \mathbb{R}^{L \times L}$.
With this attention matrix, the attention weight matrix is defined as below,
\begin{equation}
A\left[i,j\right] = \frac{QK^\top}{\sqrt{D}} M\left[i,j\right],
\label{eq:A_space}
\end{equation}
where $M\left[i,j\right]$ is multiplied with the attention weight in (\ref{eq:A}).
If $M$ is very sparse, in a real implementation, the computation of  (\ref{eq:A_space}) only has to consider those elements which are not masked, so the computational speed can be greatly increased. 
There are two different masks proposed in~\cite{child2019generating}, which are shown in Figure~\ref{fig:attention_strided} and Figure~\ref{fig:attention_fixed}.

\begin{figure}[t]
\begin{subfigure}{.32\columnwidth}
  \centering
  \includegraphics[width=.9\linewidth]{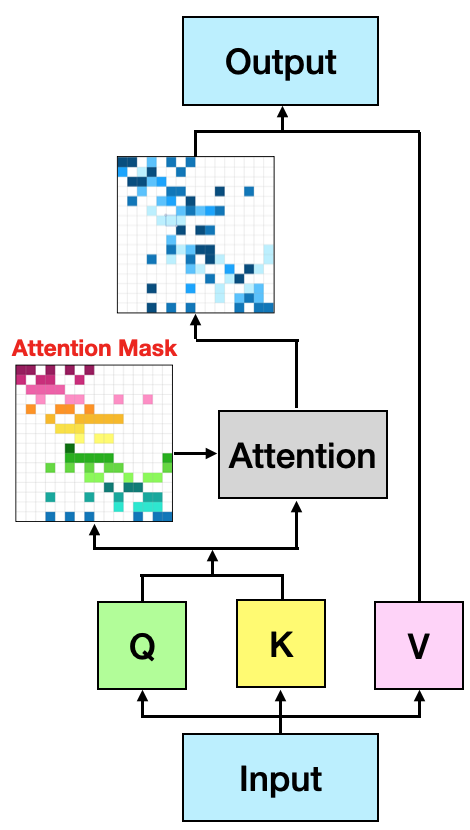}  
  \caption{Transformer}
  \label{fig:syn_transformer}
\end{subfigure}
\begin{subfigure}{.32\columnwidth}
  \includegraphics[width=.9\linewidth]{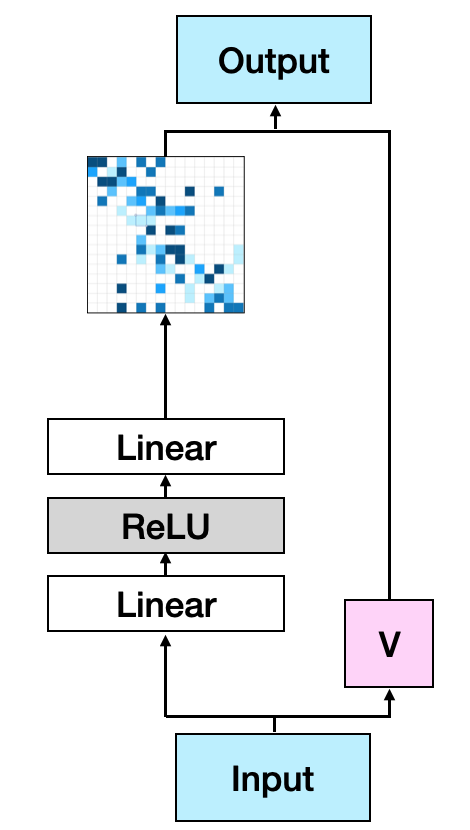}  
  \caption{Dense}
  \label{fig:syn_dense}
\end{subfigure}
\begin{subfigure}{.32\columnwidth}
  \includegraphics[width=.9\linewidth]{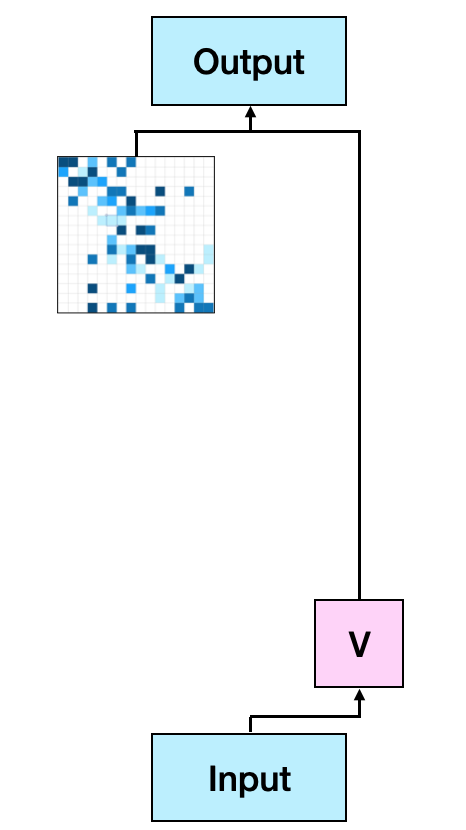}  
  \caption{Random}
  \label{fig:syn_random}
\end{subfigure}
\caption{Comparison between transformers and SYNTHESIZER}
\label{fig:SYNTHESIZER}
\end{figure}

\subsection{SYNTHESIZER}
Instead of learning attention masks by algorithms, SYNTHESIZER~\cite{tay2020SYNTHESIZER} learns the attention weights directly. 
We compare the typical attention weight generation process of Transformer and SYNTHESIZER in Figure~\ref{fig:SYNTHESIZER}.
In Figure~\ref{fig:syn_transformer}, typical transformer-based models use queries (Q) and keys (K) to generate attention weights, while in Figure~\ref{fig:syn_dense} and Figure~\ref{fig:syn_random}, SYNTHESIZER directly learns attention weights directly. In addition, although queries (Q) and keys (K) are removed in SYNTHESIZER, values (V) should be kept here.
We implement both versions of SYNTHESIZER in this paper.

\subsubsection{Dense SYNTHESIZER}
\label{sec:dense_syn}
In Figure~\ref{fig:syn_dense}, the attention weights
$A$
are generated by feeding the input
$X$
to a function 
$F$
with two hidden layers, as the left flow in Figure~\ref{fig:syn_dense}. 
Here, $W_{1}:\mathbb{R}^D\rightarrow\mathbb{R}^N$ and $W_{2}:\mathbb{R}^N\rightarrow\mathbb{R}^L$:
\begin{align}
    F\left(X\right)=W_2\left(\sigma_R\left(W_1\left(X\right)+b_1\right)\right)+b_2,
\end{align}
where $\sigma_R$ is the ReLU activation function and 
$N$
is a user-defined hyperparameter.

\subsubsection{Random SYNTHESIZER}
\label{sec:random_syn}
In Figure~\ref{fig:syn_random}, we define $A\ \left (A\in H\times L\times L\right)$ a learnable matrix\footnote{In other words, the elements in $A$ are considered as the network parameters.}, which is a randomly initialized matrix, learned with the other parts of the network.
Note that $A$ does not depend on the input sequences. It is the same across all the inputs.
Namely, data in a batch share the same attention weights.

\subsection{Our method}
Our proposed model architecture is mainly based on Random SYNTHESIZER, the attention weights of which are input-independent and fixed after pre-training. We use MAM to pre-train our models on a great amount of unlabeled data. 

Another major issue of our model is the initialization policy of attention weights. Our experimental results reveal that how we initialize those attention weights do have great influence on all downstream tasks. 

As a result, we try to analyze those attention weights of baseline models so as to gain an in-depth and holistic understanding of the attention mechanism in audio transformer-based models. We aimed for analyzing and visualizing the attention weights of baseline models with t-SNE~\cite{maaten2008visualizing} and PCA~\cite{shlens2014tutorial}, while also attempting to identify some general trends and significant patterns among them. Our ultimate goal is to make our model imitate the behavior of the typical transformer-based models yet using less time in both training and inference.

\begin{figure}[t]
\centering
\begin{subfigure}[t]{.24\columnwidth}
  \includegraphics[width=.99\textwidth]{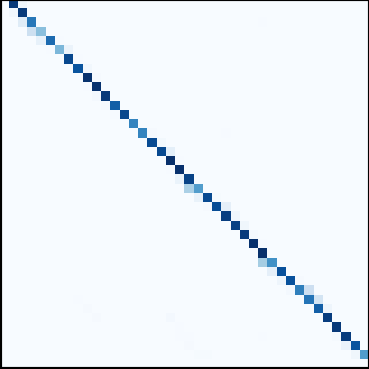}
  \caption{Diagonal}
  \label{fig:our_diag}
\end{subfigure}
\begin{subfigure}[t]{.24\columnwidth}
  \includegraphics[width=.99\textwidth]{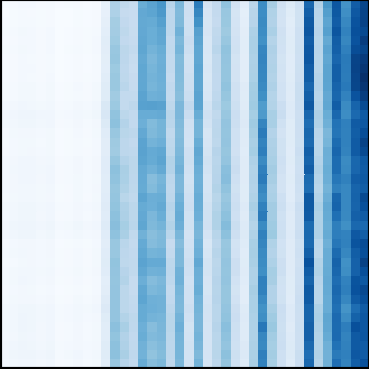}  
  \caption{Increasing)}
  \label{fig:our_incre}
\end{subfigure}
\begin{subfigure}[t]{.24\columnwidth}
  \includegraphics[width=.99\linewidth]{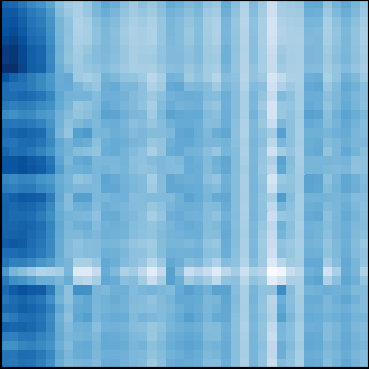}  
  \caption{Decreasing}
  \label{fig:our_decre}
\end{subfigure}
\begin{subfigure}[t]{.24\columnwidth}

  \includegraphics[width=.99\linewidth]{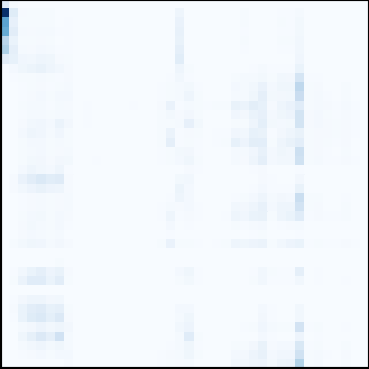}  
  \caption{Sparse}
  \label{fig:our_sparse}
\end{subfigure}

\caption{Four kinds of common attention weights in pre-trained baseline models.}
\label{fig:our_method}
\end{figure}

Figure~\ref{fig:our_method} shows the most common attention weights in the pre-trained baseline model. Basically, all attention weights can be categorized into these four cases. All of them represent a distinct relation between Q and K respectively. Moreover, along with the visualization of PCA and t-SNE (we will show in Section~\ref{sec:tsne_pca}), we use the \textit{fixed attention weights} shown in Figure~\ref{fig:attenion_weight_init} to initialize our model. Specifically, in Figure~\ref{fig:attenion_weight_init}, the five leftmost attention weights correspond to the \textit{Diagonal} case, the five rightmost ones correspond to the \textit{Sparse} case, and the middle two weights correspond to the \textit{Increasing} and \textit{Decreasing} cases.
With this decent initialization, our model can achieve comparable performance to the baseline models.

\renewcommand{\thefootnote}{\fnsymbol{footnote}}
\newcommand{\tablefootnotemark}[1]{\textsuperscript{\getrefnumber{#1}}}

\begin{table*}[!htbp]
  \caption{Summary of all attentions. $L$: input sequence length, $D$: hidden dimension, $H$: number of attention heads, $C$: number of keys attended to, and $N$: hidden dimension in SYNTHESIZER (Dense). The unit of column \textbf{Practical Time} is sec/1000 batches.}
  \label{tab:summary}
  \centering
  \resizebox{0.85\textwidth}{!}
{
  \begin{tabular}{l|cccc}
    \toprule
    \textbf{Attention}      & \textbf{Theoretical Time (Training)}   &\textbf{Theoretical Time (Inference)}   & \textbf{Practical Time (Inference)} & \textbf{Application Fields}  \\
    \midrule
    Baseline (QK)~\cite{vaswani2017attention}            
    & $4LD+2HL^{2}$ 
    &  $2LD+HL^{2}$  
    &  146.8
    & Speech,CV,NLP       \\
    
    Baseline (Q)~\cite{kitaev2020reformer}             
    & $2LD+2HL^{2}$    
    & $LD+HL^{2}$      
    & 143.4
    & CV,NLP            \\
    
    \midrule
    \rowcolor{yellow!40}
    Sparse (strided)~\cite{child2019generating}         
    & $2LD+2HL\sqrt{L}+HL\sqrt{L}$    
    &  $LD+HL\sqrt{L}+HL\sqrt{L}/2$  
    &  143.4
    &CV                  \\
    \rowcolor{yellow!40}
    Sparse (fixed)~\cite{child2019generating}            
    & $2LD+HL\sqrt{L}+HL\sqrt{L}$    
    & $LD+HL\sqrt{L}/2+HL\sqrt{L}/2$     
    & 143.4
    &CV                \\
    
    \rowcolor{LightCyan}
    Sign-ALSH~\cite{shrivastava2014improved}                
    & $2LD+(HL+2HL^{2})/2+2HLC$ 
    & $LD+(HL+2HL^{2})/2+HLC$                
    & 268.9
    & RS\tablefootnote{Recommender Systems\label{fn1}}
    \\
    
    \rowcolor{LightCyan}
    XBOX~\cite{bachrach2014speeding}                     
    & $2LD+(HL+2HL^{2})/2+2HLC$    
    & $LD+(HL+2HL^{2})/2+HLC$   
    & 235.5
    & RS                \\
    
    \rowcolor{LightCyan}
    XBOX (QNF)~\cite{huang2018accurate}               
    & $2LD+(HL+2HL^{2})/2+2HLC$    
    & $LD+(HL+2HL^{2})/2+HLC$  
    & 240.2
    & RS                \\
    
    \rowcolor{LightCyan}
    Simple LSH~\cite{neyshabur2014symmetric}               
    & $2LD+(HL+2HL^{2})/2+2HLC$    
    & $LD+(HL+2HL^{2})/2+HLC$   
    & 236.2
    & RS                \\
    
    \rowcolor{LightCyan}
    Simple ALSH~\cite{neyshabur2014symmetric}                 
    & $2LD+(HL+2HL^{2})/2+2HLC$    
    & $LD+(HL+2HL^{2})/2+HLC$                      
    & 235.7
    & RS              \\
    
    \rowcolor{green!10}
    SYN. (Dense)~\cite{tay2020SYNTHESIZER}      
    & $2LN+2L^{2}$    
    & $LN+L^{2}$              
    & 111.0
    & NLP                \\
    
    \rowcolor{green!10}
    SYN. (Dense+M\tablefootnote{M denotes multi-head attention.\label{fn:mh}}
)~\cite{tay2020SYNTHESIZER}      
    & $2HLN+2HL^{2}$    
    & $HLN+HL^{2}$      
    & 130.5
    & NLP                \\
    
    \rowcolor{green!10}
    SYN. (Random)~\cite{tay2020SYNTHESIZER}     
    & $HL^{2}$    
    & -              
    & 116.8
    & NLP                \\
    
    \rowcolor{green!20}
    Ours       
    & $HL^{2}$    
    & -                    
    & 116.8
    & -                \\
    \bottomrule
  \end{tabular}
}
\end{table*}

\begin{table}[ht]
  \caption{Performance of all attentions}
  \label{tab:performance}
  \centering
  \resizebox{0.85\columnwidth}{!}{
  \begin{tabular}{l|cccc}
    \toprule
    \multirow{2}{*}{\textbf{Attention}} &\multicolumn{2}{c}{\textbf{Speaker}} &\multicolumn{2}{c}{\textbf{Phoneme}} \\
    &\textbf{Utterance} &\textbf{Frame} &\textbf{1-hidden} &\textbf{2-hidden}\\
    \midrule
    Baseline (Mel\tablefootnote{Apply input acoustic features (Mel features) directly to downstream models.\label{fn:mel}})          &0.0060     &0.0033      &0.5246    &0.5768         \\
    Baseline (QK)           &0.9926    &0.9824      &0.6460    &0.6887         \\
    Baseline (Q)             &0.9898    &0.9622      &0.5893    &0.6345         \\
    \midrule
    \rowcolor{yellow!40}
    Sparse (Strided)         &0.9786    &0.9039      &0.6048    &0.6450         \\
    \rowcolor{yellow!40}
    Sparse (Fixed)            &0.9597    &0.7960      &0.6069    &0.6846         \\
    \rowcolor{LightCyan}
    Sign-ALSH                &0.9716    &0.8237      &0.5863    &0.6393         \\
    \rowcolor{LightCyan}
    XBOX                     &0.9639    &0.7994      &0.5860    &0.6262         \\
    \rowcolor{LightCyan}
    XBOX (QNF)               &0.9667    &0.7958      &0.5819    &0.6241         \\
    \rowcolor{LightCyan}
    Simple LSH               &0.9628    &0.7370      &0.5771    &0.6189         \\
    \rowcolor{LightCyan}
    Simple ALSH              &0.9678    &0.7999      &0.5783    &0.6214         \\
    \rowcolor{green!10}
    SYN. (Dense)      &0.9660    &0.9027      &0.6180    &0.6287         \\
    \rowcolor{green!10}
    SYN. (Dense+M\tablefootnotemark{fn:mh})    &0.9509    &0.9135      &0.6073    &0.6471         \\
    \rowcolor{green!10}
    SYN. (Random)     &0.9803    &0.8868      &0.5820    &0.6237         \\ 
    \rowcolor{green!20}
    Ours       &0.9842    &0.9855      &0.6157    &0.6492         \\
    \bottomrule
  \end{tabular} 
  }
\end{table}

\section{Experiments}
All attention algorithms and SYNTHESIZER models are compared in Table~\ref{tab:summary}. There are three main groups, which stand for Sparse Attention, LSH, and SYNTHESIZER respectively. We compare the theoretical time complexity of both training and inference and the practical running time; we also list the corresponding application fields where each attention has been applied. Specifically, the practical running time here is the sum of $1000$ training examples whose sequence length equals to 500.

Here, we can make seven key observations:
\begin{inparaenum}[1)]
  \item Most attention algorithms have not been applied to many application fields, showing that further research may be necessary.
  \item Sparse transformer has similar practical running time to those of the typical transformer-based models.
  \item Although the theoretical time complexity shows the benefit when $L$ is quite large, these LSH/ALSH algorithms do not actually accelerate inference speed. Even for those much longer sequences (e.g.,1000), their inference speed still does not outperform the baseline models (Due to the space limitations, we do not list the results in this table).
  \item All the second terms of LSH/ALSH are for hashing and we use 16-bit floating points to implement. This is why we multiply $1/2$ here. Also, there is no gradient during hashing; thus, the training time complexity is the same as inference complexity.
  \item Our models and all kinds of Syn. indeed accelerate both training and inference. 
  \item For Syn. (Dense) and Syn. (Dense+M), letting $N\ll L$ can accelerate inference dramatically; we let $N=16$ in this work.
  \item Our models and Syn. (Random) do not need any computation to generate attention weights during inference, i.e., their attention weights are both input-independent and totally fixed after pre-training.
\end{inparaenum}

\begin{figure}[t]
\centering
\frame{\includegraphics[width=.99\linewidth]{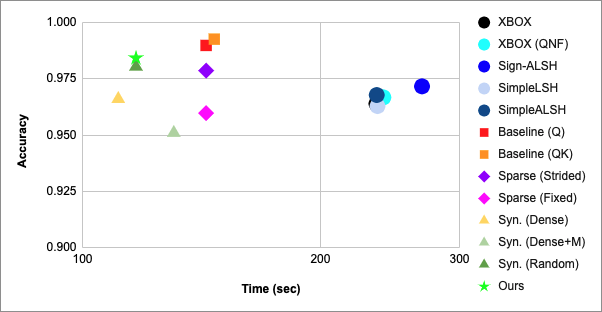}}

\caption{Performance of utterance-level speaker classification versus practical running time.}
\label{fig:performance_time}
\end{figure}

\subsection{Pre-training}

We pre-train the transformer-based models on the unlabeled audio data (Librispeech~\cite{panayotov2015librispeech} \textit{train-clean-360} subset) and fine-tune for the downstream tasks. 
All transformer-based models have the same architecture as a 6-layered Audio ALBERT (AALBERT)~\cite{chi2019aalbert}.
We adopt the shared-QK attention~\cite{kitaev2020reformer} for all the methods mentioned in Sections~\ref{sec:LSH} and Sections~\ref{subsec:sparse}, which ties the weights of Q and K to further reduce the computation requirement. 
All the other settings of the pre-training stage follow those of ~\cite{chi2019aalbert}. 
The models are trained for 500k steps with a batch size of 50 and a learning rate of 5e-5. The optimizer is LAMB~\cite{you2019large}. Lastly, for the two hyperparameters in Section~\ref{sec:sign-alsh}, we let $U=0.75$ and $m=2$, which is shown to have the best empirical result~\cite{shrivastava2014improved}.


\subsection{Performance of downstream tasks}
We evaluate the self-supervised models on three downstream tasks, including utterance-level speaker classification (on the \textit{train-clean-100} subset), frame-level speaker classification (on the \textit{train-clean-100}), and phoneme classification (on the \textit{train-clean-360} subset with phoneme labels).  
In downstream tasks, the pre-trained models are used as feature-extractors whose parameters are fixed during fine-tuning.
For speaker classification, the extracted representations in the utterance-level task are passed to an average pooling layer, then the mean representation is fed to a linear classifier. As in frame-level task, we simply train a linear classifier to predict the speaker label of every audio frame.
As for the phoneme classification, we utilize both a one-hidden-layer model and a two-hidden-layer model in downstream tasks; a layer consists of a linear layer along with a ReLU activation function here. 

The results are shown in Table~\ref{tab:performance}.
Baseline (QK) and Baseline (Q) (shared-QK attention) remarkably outperform Baseline (Mel), which shows the importance of pre-training. 
LSH/ALSH algorithms have negative influences on most downstream tasks, 
showing that restricting the attention by LSH/ALSH algorithm is not effective enough. 
For utterance-level speaker classification, the average pooling layer in the downstream model acts like a global attention mechanism, which compensates the effects of LSH/ALSH and thus have better performance than those of frame-level. 
Sparse Attention obtains higher accuracy than LSH/ALSH, which shows that local information might be important since Sparse Attention always contains a fixed-size local window whereas LSH/ALSH do not. 
SYNTHESIZER models perform even better on average than the others; however, they fail to match Baseline (Q) on the frame-level speaker classification task.
Our model, which is based on SYNTHESIZER with some specific attention weights initialization, achieves competitive performances compared to Baseline (Q) and outperforms Baseline (QK) on the frame-level speaker classification task. 

Finally yet importantly, Figure~\ref{fig:performance_time} shows the accuracy of utterance-level speaker classification versus the practical running time. The models in the upper left corner are the best ones; they have higher accuracy as well as faster training/inference speed.
This figure shows that the proposed approach achieves better performance with faster speed.

\subsection{t-SNE and PCA}
\label{sec:tsne_pca}

\begin{figure}[t]
\centering
\frame{\includegraphics[width=.8\linewidth]{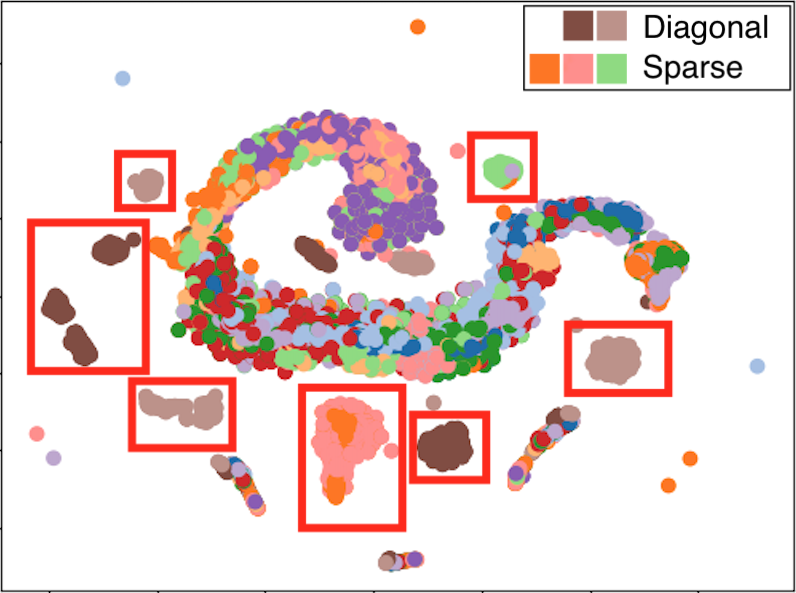}}
\caption{Visualization of attention weights in the baseline model.}
\label{fig:tsne_pca}
\end{figure}

In addition to observing the attention weights in baseline models, such as those in Figure~\ref{fig:our_method}, we also visualize them with t-SNE and PCA. We first reshape $2$-d attention weights matrices to $1$-d ones. For instance, a $2$-d matrix $A \in R^{500\times 500}$ will be reshaped as a $1$-d matrix $A^\prime \in R^{1\times250000}$. Then we use PCA to reduce their dimensionality to $900$. Subsequently, we reduce their dimensionality to $2$ with t-SNE. Figure~\ref{fig:tsne_pca} shows the result; 12 different colors represent 12 attention heads. Among all, two heads generate \textit{Diagonal} attention weights, three heads usually generate \textit{Sparse} attention weights, while the other heads output some patterns with vertical stripes, some of which are increasing and some of which are decreasing.
Ultimately, Figure~\ref{fig:attenion_weight} shows the attention weights of our model after pre-training. They are very similar to those in Figure~\ref{fig:attenion_weight_init}, showing that this set of attention weights may be expressive enough to capture the information of data.

\section{Conclusion}
We explore the possibility of reducing computation complexity in audio transformer-based models for self-supervised representation learning. We try LSH/ALSH and Sparse attention to limit the number of token-token interaction in self-attention. We also introduce recently proposed attention modules SYNTHESIZER. Then, we propose to combine SYNTHESIZER with some special initialization. In the experiments, our proposed architecture not only performs comparably to typical transformer-based models on downstream tasks but requires almost $20\%$ less training/inference time and similar memory usage.

\bibliographystyle{IEEEbib}
\bibliography{strings,refs}

\end{document}